\documentclass[twocolumn]{aastex631}

\shorttitle{Black hole coupling \& Accretion History}
\shortauthors{Lacy, et al.}

\begin{document}

\title{Constraints on cosmological coupling from the accretion history of supermassive black holes}

\correspondingauthor{Mark Lacy}
\email{mlacy@nrao.edu }

\author[0000-0002-3032-1783]{Mark Lacy}
\affiliation{National Radio Astronomy Observatory, Charlottesville, VA, USA}

\author[0000-0001-6970-7782]{Athena Engholm}
\affiliation{Department of Physics and Astronomy, University of Hawai‘i at M\~{a}noa, 2505 Correa Rd., Honolulu, HI 96822, USA}
\affiliation{Institute for Astronomy, University of Hawai‘i, 2680 Woodlawn Dr., Honolulu, HI 96822, USA}

\author[0000-0003-1748-2010]{Duncan Farrah}
\affiliation{Department of Physics and Astronomy, University of Hawai‘i at M\~{a}noa, 2505 Correa Rd., Honolulu, HI 96822, USA}
\affiliation{Institute for Astronomy, University of Hawai‘i, 2680 Woodlawn Dr., Honolulu, HI 96822, USA}

\author[0009-0002-6970-5247]{Kiana Ejercito}
\affiliation{Department of Physics and Astronomy, University of Hawai‘i at M\~{a}noa, 2505 Correa Rd., Honolulu, HI 96822, USA}
\affiliation{Institute for Astronomy, University of Hawai‘i, 2680 Woodlawn Dr., Honolulu, HI 96822, USA}

\begin{abstract}
Coupling of black hole mass to the cosmic expansion has been suggested as a possible path to understanding the dark energy content of the Universe. We test this hypothesis by comparing the supermassive black hole (SMBH) mass density at $z=0$ to the total mass accreted in AGN since $z=6$, to constrain how much of the SMBH mass density can arise from cosmologically-coupled growth, as opposed to growth by accretion. Using an estimate of the local SMBH mass density of $\approx 1.0\times10^{6}\,$M$_{\odot}\,$Mpc$^{-1}$, a radiative accretion efficiency, $\eta$: $0.05<\eta<0.3$, and the observed AGN luminosity density at $z\approx 4$, we constrain the value of the  coupling constant between the scale size of the Universe and the black hole mass, $k$, to lie in the range $0<k\stackrel{<}{_{\sim}}2$, below the value of $k=3$ needed for black holes to be the source term for dark energy. 
Initial estimates of the gravitational wave background using pulsar timing arrays, however, favor a higher SMBH mass density at $z=0$. We show that if we adopt such a mass density at $z=0$ of $\approx 7.4\times 10^{6}\,$M$_{\odot}\,$Mpc$^{-1}$, this makes $k=3$ viable even for low radiative efficiencies, and may exclude non-zero cosmological coupling. We conclude that, although current estimates of the SMBH mass density based on the black hole mass -- bulge mass relation probably exclude $k=3$, the possibility remains open that, if the GWB is due to SMBH mergers, $k>2$ is preferred.

\end{abstract}


\section{Introduction} \label{sec:intro}

\subsection{The Soltan Argument}
There exists an elegant argument for constraining the growth of supermassive black hole (SMBH) mass in galaxies via accretion. \citet{1982MNRAS.200..115S} showed that the integral of the AGN luminosity density over cosmic history gives an estimate of the total accreted mass onto SMBHs, which can be compared to the local mass density of SMBHs to constrain how they grow. This relation, between two independently derived quantities, can give excellent constraints on both AGN physics and the nature of SMBHs themselves.  For example, using then available data, \citet{1982MNRAS.200..115S} showed that the history of accreted mass is close to the local SMBH mass density, if a radiative efficiency of $\eta\approx 0.1$ is assumed.  This range in $\eta$ is in good agreement with other estimates. Theoretical arguments give a range of $0.05\lesssim\eta\lesssim0.43$, depending on the spin of the black hole (BH) \citep{1970Natur.226...64B,1974ApJ...191..507T}. This range is itself in good agreement with most other studies, which find values in the range $0.1\lesssim\eta\lesssim0.3$ \citep[e.g.][]{2002MNRAS.335..965Y,2002ApJ...565L..75E,2007ApJ...654..731H, 2009ApJ...690...20S,2015ApJ...802..102L,2017SCPMA..60j9511Z,2020MNRAS.493.1500S, 2020MNRAS.495.3252S,farrah22}. This consistency argues for the increase in SMBH mass density with cosmic time to be driven by accretion. 

\subsection{Cosmologically-coupled black holes}
 
For several decades it has been recognized that BHs with non-singular interiors (including `dark energy'-like interiors) may have exteriors identical to BHs with interior singularities \citep[e.g.][]{1966JETP...22..378G,1992GReGr..24..235D,farja07, 2019PhRvD..99d4037B,2023arXiv230916444C}. Recently, however, it was realized that the masses of non-singular BHs could be affected by the cosmic expansion \citep{cw19}. Indeed, in a recent paper, \citet{2023JCAP...11..007C} show that such a coupling is necessary in General Relativity (GR) if BHs do not contain singularities. Thus the detection of a non-zero cosmological coupling would have important implications for our understanding of BHs.

The mass coupling between BHs and the cosmic expansion can be parameterized via:
\begin{equation}\label{eqn:defk}
M_{\bullet}(a) = M_{\bullet}(a_i)\left(\frac{a}{a_{i}}\right)^k 
\end{equation}
\noindent \citep{cnf2020,2021ApJ...921L..22C} in which $M_{\bullet}(a)$ and $M_{\bullet}(a_i)$ are the BH mass at some later and earlier times respectively, $a$ and $a_{i}$ are the scale factors at those times ($a=(1+z)^{-1}$ at redshift $z$), and $k$ is the cosmological coupling strength, with $-3\leq k \leq 3$ from a causality constraint. 

\citet{cw19} show that the value of $k$ is related to the equation of state of the medium internal to the BH. Expressing the equation of state as $p=w\rho$, where $p$ is the pressure and $\rho$ the density, \citet{2021ApJ...921L..22C} show that $k=-3 w$. Thus, for vacuum energy where $p=-\rho$, $k=3$.
In this case, the mass density of BHs is conserved as the Universe expands: their number density decreases with $a^{-3}$ but their mass increases as $a^3$ and so BHs could account for the cosmological constant \citep{cw19,2023ApJ...944L..31F} \citep[see also ][for an alternative way BHs could be the origin of dark energy]{2009PhRvD..80d3513P}. In contrast, GR solutions for non-singular BHs prefer $k=1$ \citep{2023JCAP...11..007C}. In this case, the BH mass growth would not compensate for cosmological dilution of their density, so they would not contribute to dark energy. 

Observationally, the evidence is mixed. \citet{farrah23ellip, 2023ApJ...944L..31F} find, based on an analysis of SMBH versus host galaxy masses out to $z\approx 2$, that a value of $k\approx 3$ is suggested and $k=0$ can be ruled out at $\approx 3\sigma$. 
On the other hand, 
\citet{2023arXiv230503408L} argue, based on a much smaller sample of AGN at $z>4.5$ with much lower stellar masses, that $k=3$ can be ruled out at $\approx 2\sigma$. For stellar mass black holes, \citet{2023ApJ...956..128G} argue that cosmological coupling of black holes can explain the distribution of compact object masses in low mass X-ray binaries, in particular the ``mass gap'' between the highest mass neutron stars and lowest mass BHs. However,  \cite{2023ApJ...947L..12R} disfavor coupling based on observations of a BH binary in the globular cluster NGC 3201 and \citet{andrae23} argue that two BH binaries found by {\em Gaia} are likely to have formed with BH masses that were too small if $k=3$.\footnote{It is worth noting that these constraints from BH binaries assume that the Tolman-Oppenheimer-Volkov limit for the minimum mass of a BH of $\sim2.2M_{\odot}$ holds.} Also, \citet{2023OJAp....6E..25G} argue that $k=3$ would result in too many BH mergers compared to observations from gravitational wave detectors.

\subsection{Scope of this paper}

As the observational picture seems far from settled, we decided to investigate an independent constraint on BH growth. For cosmologically-coupled mass growth of BHs to be viable, the mass increase must fit within the constraint provided by the Soltan argument - that is, the local SMBH mass density must accommodate both the mass increase due to accretion, and the mass increase due to cosmological coupling.  In this paper, we examine whether cosmologically-coupled BH growth can be accommodated within the constraints from the Soltan argument, and, if so, what constraints this can set on the value of $k$. We adopt the latest measures of both the AGN luminosity function (LF) over cosmic history, and of the local SMBH mass density, and explore values for the cosmological coupling strength in the range $0<k<3$. We discuss uncertainties on our results arising from uncertainties in both $\eta$ and the Eddington ratio $\lambda$. \S\ref{sec:meth} presents our methods and \S\ref{sec:data} our sources of data. \S\ref{sec:res} presents our results and  \S\ref{sec:disc}  discusses these results in the context of uncertainties on observed parameters. \S\ref{sec:conc} summarizes our conclusions. We assume a cosmology with $H_0=70\, {\rm kms^{-1}Mpc}$, $\Omega_{\rm M}=0.3$ and $\Omega_{\Lambda}=0.7$.

\section{Methods}\label{sec:meth}

\subsection{Generalization of the Soltan Argument for cosmologically-coupled black holes}\label{sec:genSoltan}

The SMBH mass and luminosity are connected for each AGN through 
\begin{equation}
L = \eta \dot{M}_{\rm acc} c^2,
\end{equation}
where $L$ is the bolometric luminosity of the AGN, $\eta$ is the radiative efficiency, and $\dot{M}_{\rm acc}$ is the mass accretion rate of the BH. Mass/energy that is not radiated is accreted onto the BH, so the rate of mass increase of the BH, $\dot{M}_{\bullet}=(1-\eta) L_{\rm{bol}}/(\eta c^2)$.

In the absence of cosmological coupling (and ignoring the mass density of any high redshift seed population, which we assume to be negligible throughout this paper, the mass density in SMBHs today is equal to the integral of the AGN luminosity density, $U(z)$ over Cosmic Time, modulated by $\eta$: 

\begin{equation}\label{eqn:soltan}
    \rho_{\bullet}(0)= \int_{0}^{\infty} \left(\frac{1-\eta}{\eta} \right) \frac{U(z)}{c^{2}} \frac{\mathrm{d} t}{\mathrm{d} z} \mathrm{d}z,
\end{equation}
where $U$:

\begin{eqnarray}
U(z)&=&\int_{\mathrm{L_{min}}}^{\mathrm{L_{max}}} L \frac{\partial \phi(L,z)}{\partial\mathrm{(log_{10}} L) }\,\mathrm{d(log_{10}} L) \\
&=&\frac{1}{\mathrm{ln} 10}  \int_{\mathrm{L_{min}}}^{\mathrm{L_{max}}} \frac{\partial \phi(L,z)}{\partial\mathrm{(log_{10}} L) }\, \mathrm{d}L,
\end{eqnarray}
$\frac{\partial \phi(L,z)}{\partial\mathrm{(log_{10}} L)}$is the bolometric luminosity function of AGN and 
\begin{equation}
    \frac{\mathrm{d} t}{\mathrm{d} z} = \frac{-1}{(1+z) H(z)},
\end{equation}  
where
\begin{equation}
    H(z) =  H_0 \sqrt{(1+z)^3 \Omega_{\mathrm m} + \Omega_{\Lambda}} \ ,
\end{equation}
and $H_0$ is the Hubble Constant at the current epoch.
If, in addition, BH mass is increasing with the scale factor of the Universe as per Equation \ref{eqn:defk}, then the contribution to the BH mass density at redshift $z'$, $\delta \rho_{\bullet}(z')$, to that at redshift $z$ is 
\begin{equation}
    \delta \rho_{\bullet}(z) (1+z)^{k}=\delta \rho_{\bullet}(z') (1+z')^k
\end{equation}

and the version of Equation \ref{eqn:soltan} with cosmologically-coupled BHs is thus:
\begin{equation}\label{eqn:generalSoltan}
    \rho_{\bullet}(z)=  \frac{1}{(1+z)^k}\int_{\infty}^{z} \left( \frac{1-\eta}{\eta} \right) \frac{U(z')(1+z
    ')^{k-1}}{H(z')c^2} \,\mathrm{d} z'
\end{equation}

\subsection{Constraints from the AGN luminosity function at high redshifts}\label{highzLF}

The high redshift AGN luminosity function also provides a constraint on SMBH growth. If the growth of SMBHs by accretion at high redshifts is too slow, there is insufficient mass density in SMBHs to power the observed AGN luminosity density at early times ($z\stackrel{>}{_{\sim}}2$), when a large fraction of the SMBH population was actively accreting. We relate the luminosity density in AGN to the mass density in SMBHs via a mean Eddington ratio, $\lambda=L/L_{\rm Edd}$, where $L_{\rm Edd}$ is the Eddington luminosity. (We note that both this analysis and the Soltan analysis above could be made more constraining by considering the observed distribution of SMBH masses and AGN luminosities and using a continuity-equation approach \citep[e.g.][] {raimundo12,2017A&A...600A..64T}. However, given the uncertainties in our understanding of the possible mass/luminosity dependencies of variables such as $\lambda$ and $\eta$, and uncertainties in the exact forms of the mass and luminosity functions and SMBH merger histories, we retain an integral approach for this study.)

\section{Data}\label{sec:data}
Two measurements are needed to use the formalism in \S\ref{sec:meth}. First is a measurement of the local SMBH mass density. Second is a measure of the accretion history of SMBHs since the first quasars, via a bolometric AGN luminosity function. We describe both these sources of data in this section. 

\subsection{The local mass density of SMBHs}\label{sec:bhdensity}

\begin{figure*}[ht]
    \includegraphics[trim= 0.5in 0.0in 0.0in 0.0in,scale=0.85]{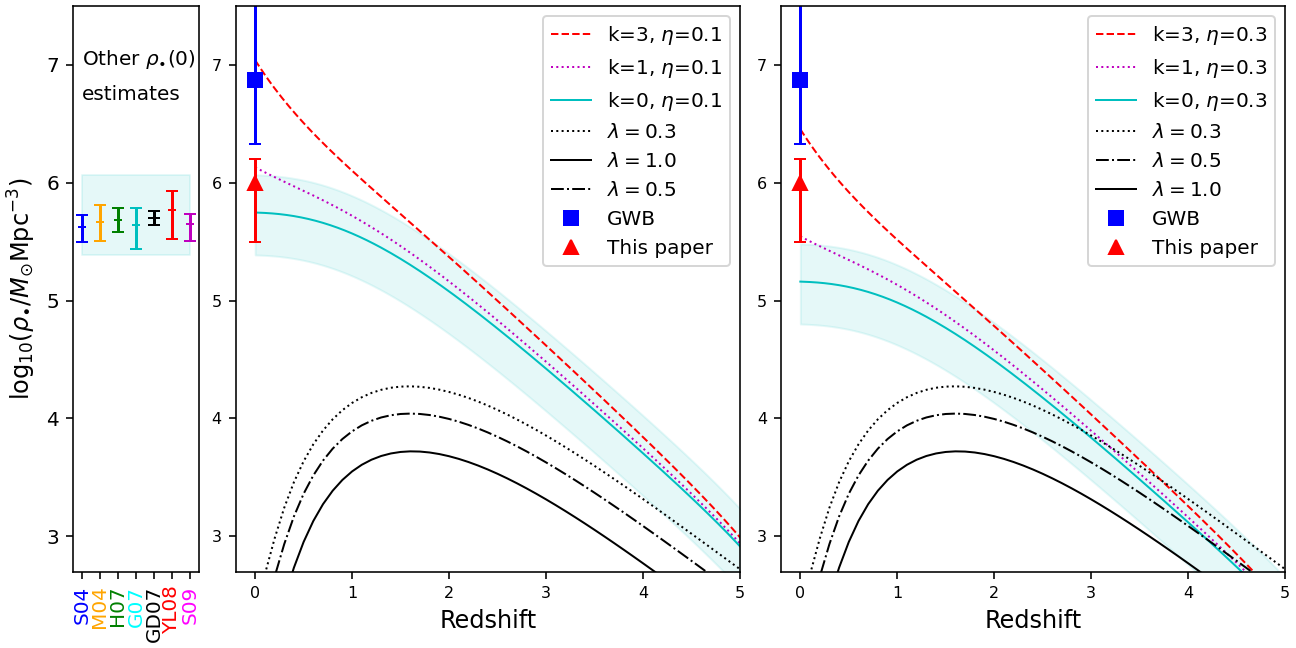}
    \caption{ {\em Middle and right panels:} constraints on the value of the cosmological coupling strength $k$ (Equation \ref{eqn:defk}) based on comparing the growth of SMBH density via accretion to the local density of SMBHs. {\em Middle:} assuming $\eta=0.1$ (which agrees well with the GD07 estimate of $\rho_{\bullet}(0)$ if $k=0$) and {\em right:} $\eta$=0.3, the maximum theoretical value for a maximally-spinning BH \citep{1974ApJ...191..507T}. In both, the red triangle and error bar represents our estimate of the local SMBH mass density from the $M_{\bullet}-M_{\rm bulge}$ relation and the blue square and error bar (which extends well above the plotted range) the preliminary estimate from the GWB, as discussed in \S 3.2. Also in both, the black lines represent the minimum mass density in SMBHs needed to account for the observed luminosity function for different mean values of $\lambda$ and the cyan-solid, magenta-dotted and red-dashed lines the growth in BH mass density for different values of $k$ assuming the AGN luminosity function of \citet{2020MNRAS.495.3252S}. The shaded region around the $k=0$ (cyan-solid) line corresponds to the uncertainty in \citet{2020MNRAS.495.3252S} (uncertainty ranges for other values of $k$ are similar). {\em Far left} a compilation of estimates of the local mass density in black holes from the literature: S04: \citet{2004MNRAS.354.1020S}, M04: \citet{2004MNRAS.351..169M}, H07: \citet{2007ApJ...654..731H}, G07: \citet{2007MNRAS.378..198G}, GD07: \citet{2007MNRAS.380L..15G}, YL08: \citet{2008ApJ...689..732Y}, S09: \citet{2009ApJ...690...20S}, with the cyan shaded area indicating the range of uncertainty at $z=0$ for the \citet{2020MNRAS.495.3252S} luminosity function, assuming $\eta=0.1$. 
    }
    \label{fig:scenario1}
\end{figure*}

\citet[hereafter GD07]{2007MNRAS.380L..15G} compared several estimates of the local SMBH mass density available in the literature at the time(see Figure \ref{fig:scenario1}, left-hand panel). These were derived from the SMBH mass --  bulge velocity dispersion ($M_{\bullet} -\sigma$) and the SMBH mass -- bulge mass  ($M_{\bullet} - M_{\rm bulge}$) relations, and adjusted to a common $H_0=70 {\rm kms^{-1} Mpc^{-1}}$. GD07 determined a local SMBH mass density in the range 4.4-5.9 $\times 10^5 M_{\odot}$Mpc$^{-3}$. 
Subsequently, \citet{2013ARA&A..51..511K} presented a recalibration of the $M_{\bullet}-\sigma$ and $M_{\bullet}-M_{\rm bulge}$ relations by excluding pseudobulges and galaxy-galaxy mergers from the fitting procedure. This resulted in an increase of the $M_{\bullet}/M_{\rm bulge}$ ratio from $\approx 0.2$\% to $\approx 0.5$\%. Thus, the SMBH mass density calculated from a stellar mass function and the \citet{2013ARA&A..51..511K} $M_{\bullet}-M_{\rm bulge}$ relation results in a local SMBH mass density substantially higher than the \citet{2007MNRAS.380L..15G} estimate (we note also that the work of \citet{2013ApJ...764..184M} also suggests a relatively high normalization, $\approx 0.3$\%, however, to be conservative, we adopt the higher \citet{2013ARA&A..51..511K} normalization). Furthermore, the difference in the shapes of the stellar mass functions for disks and bulges means that the assumption of a single bulge fraction independent of stellar mass is not strictly correct.

We therefore construct an updated range for the local SMBH mass density as follows. We use the study of \citet{2016MNRAS.459...44T}, who split the stellar mass function into spheroid and disk components. Using their spheroid mass function and the $M_{\bullet} - M_{\rm bulge}$ relation of \citet{2013ARA&A..51..511K} results in an estimate of $\rho_{\bullet}(0)=1.0\times 10^6 M_{\odot}$Mpc$^{-3}$, higher than that of
\citet{2007MNRAS.380L..15G}, but lower than that obtained by assuming a fixed bulge fraction and using an all-galaxy stellar mass function.  

The uncertainty in this estimate is substantial. It may be too low -- there is evidence that a significant fraction of SMBH mass may be missed by using spheroid mass of velocity dispersion as a proxy. For example, \citet{voggel19} argue for the existence of a population of `stripped' galaxy nuclei in the haloes of giant ellipticals. These objects have modest stellar masses, but can harbor SMBHs of up to $\sim10^{7}$M$_{\odot}$. \citet{voggel19} further argue that this population may increase the total SMBH mass density by up to $\sim30\%$. At high velocity dispersions (or equivalently, high stellar masses) the $M_{\bullet}-\sigma$ relation may also be biased low, as there are several examples of `overmassive' SMBHs for their hosts \citep[e.g.][]{dullo21,night23}. Another plausible source of error is ejected SMBHs \citep{postman12,chubol22}.  Conversely, it is possible our estimate is too high. Selection biases \citep{schulze11,2016MNRAS.460.3119S} could lead to an overestimate of the SMBH mass density by a factor $\approx 3$, though the existence of this bias has not been confirmed observationally. Also, the spheroid mass function still includes pseudobulges, which, as \citet{2013ARA&A..51..511K} emphasize, typically fall below the  $M_{\bullet}-M_{\rm bulge}$ relation  \citep[though pseudobulges are typically less massive than classical bulges, making up $\stackrel{<}{_{\sim}}10$\% of the stellar mass density in spheroids]{2008ApJS..175..356H}. It is beyond the scope of this paper to synthesize these disparate results into a firm range for $\rho_{\bullet}$. However, a maximum value of  $\approx 1.6\times10^{6}$M$_{\odot}$Mpc$^{-1}$ does not appear to be ruled out (assuming $\approx 30$\% of the mass density is stripped and $\approx 30$\% ejected), and a minimum value of $3\times10^{5}$M$_{\odot}$Mpc$^{-1}$ also seems plausible, based on the most extreme estimates of bias. We therefore adopt an estimate and range of ${\rm log_{10}}(\rho_{\bullet}(0) /$M$_{\odot}$Mpc$^{-1})=6.0^{+0.2}_{-0.5}$ based on the $M_{\bullet}-M_{\rm bulge}$ relation.

Finally, we note that the recent detection of a gravitational wave background (GWB) by pulsar timing arrays, if ascribed purely to the mergers of SMBHs, results in a much higher SMBH mass density estimate than those discussed above. Using the results of \citet{nanograv23} we find that the peaks of their posterior distributions of the most relevant parameters from the fits based on their {\em Phenom+Astro} priors
(the galaxy stellar mass function characteristic density and cutoff mass, the $M_{\bullet}/M_{\rm bulge}$ ratio, and the logarithmic width of the $M_{\bullet}/M_{\rm bulge}$ distribution) lead to an estimate of $7.4\times 10^6 M_{\odot}$Mpc$^{-3}$. Using the posterior distribution of $\rho_{\bullet}$ constructed from their Monte-Carlo simulations results in a 95\% confidence range of $2-1400 \times 10^6 M_{\odot}$Mpc$^{-3}$ and an average of $\approx 1\times 10^8 M_{\odot}$Mpc$^{-3}$ \citep[we note also that][found a similar result using the earlier 12.5 yr NanoGrav data release]{2022ApJ...924...93C}. Given the large uncertainties in the SMBH merger estimates it is probably too early to be concerned with the tension between the GWB versus the $M_{\bullet} - M_{\rm bulge}$ relation \citep[see][for a discussion of the strength of the tension between the SMBH binary model and the measured GWB in the context of new physics]{2023ApJ...951L..11A}, nevertheless we consider the implications of a GWB-based estimate of $\rho_{\bullet}(0)=7.4\times 10^6 M_{\odot}$Mpc$^{-3}$ in our analysis as well.

\subsection{The bolometric AGN luminosity function}\label{sec:LF}

To compute the comoving AGN luminosity density, we need an estimate of the bolometric AGN luminosity function as a function of redshift. Integrating under this luminosity function as a function of redshift then gives the comoving AGN luminosity density.  To perform this calculation, we adopt the \citet{2020MNRAS.495.3252S} (their ``global fit B'') luminosity function (Equation \ref{eqn:shenLF}), which agrees well with, but is somewhat more refined than other estimates \citep[e.g.][]{2007ApJ...654..731H,2015ApJ...802..102L,runburg22}:

\begin{equation}\label{eqn:shenLF}
\frac{\partial \phi(L,z)}{\partial\mathrm{(log_{10}}L)}=\frac{\phi(z)^{*}}{((L/L(z)^*)^{\gamma_1(z)}+(L/L(z)^*)^{\gamma_2(z)})},
\end{equation}
where the redshift dependencies of $\phi^{*}$, $L^*$, $\gamma_1$ and $\gamma_2$ are detailed in \citet{2020MNRAS.495.3252S}.

\begin{figure*}
    \includegraphics[trim= 0.3in 0.0in 0.0in 0.0in,scale=0.90]{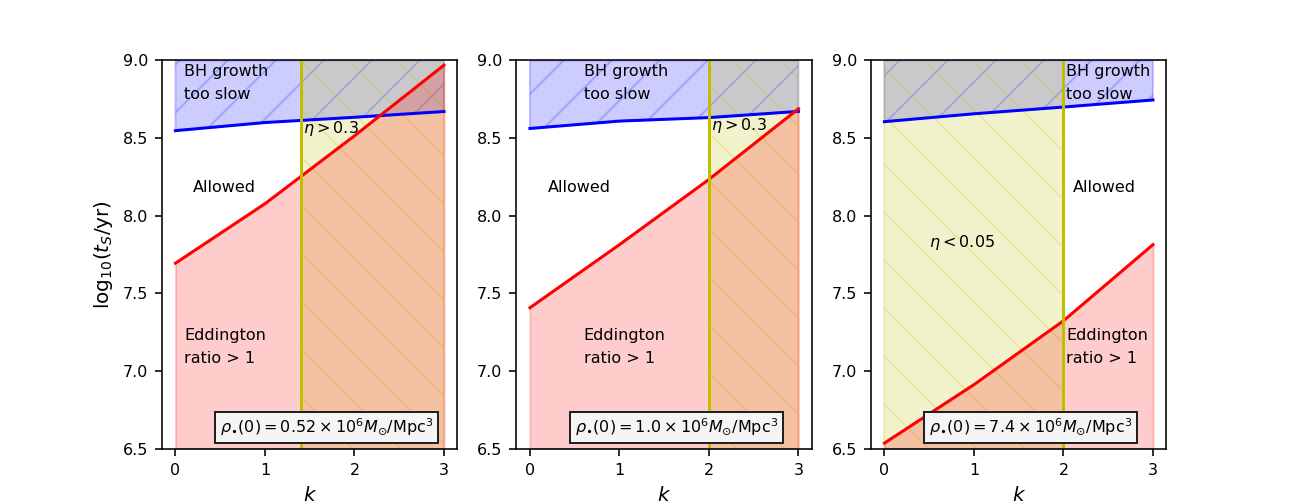}
    \caption{The allowed values of the cosmological coupling strength $k$ (Equation \ref{eqn:defk}) and the Salpeter time (Equation \ref{salptime}) $t_S$ for quasars assuming accretion-dominated growth of the supermassive black hole density. {\em Left:} assuming the GD07 estimate of $\rho_{\bullet}(0)$, {\em middle} assuming the best estimate from this paper and {\em right} assuming the value from the peaks of the GWB posterior fits. The hatched blue regions represent the parameter space where the Salpeter time is too long to grow enough black holes at $z=4$ to explain the observed luminosity function and the red regions are those where the Eddington ratio needs to exceed unity to do this.The yellow hatched regions are excluded based on the plausible range of $0.05<\eta<0.3$.}
    \label{fig:salpeter}
\end{figure*}

\section{Results}\label{sec:res}
Figure \ref{fig:scenario1} (middle and right panels) plots the SMBH mass densities from \S\ref{sec:bhdensity} with the result of integrating Equation \ref{eqn:generalSoltan} from $z=6$ to $z=0$ for fixed $\eta$ values of $0.1$ and $0.3$ \citep{1974ApJ...191..507T}. We show curves for $k=0$ (no cosmological coupling), $k=1$ (e.g. the exact solutions of \citealt{farja07}) and $k=3$ \citep{2023ApJ...944L..31F}. This indicates that, if we assume the GD07 value for the local SMBH mass density and $\langle\eta\rangle\simeq0.3$, then an upper limit can be set of $k\stackrel{<}{_{\sim}}2$. Alternatively, using the GWB value for the local SMBH mass density, then a non-zero value of $k$ is required for any reasonable value of $\eta$.

To generalize this argument, we note that, for a given value of $k$, $\eta$ can uniquely determined by matching the observed local black hole mass density, so if we assume a value of the Eddington ratio, $\lambda$ between the minimum value needed to explain the AGN luminosity function at high redshift (assuming all SMBHs are actively-accreting) and $\lambda=1$ then we can define an allowed region in AGN accretion parameter space as a function of $k$. In Figure \ref{fig:salpeter}, we use the Salpeter time:

\begin{equation}\label{salptime}
t_S=4\times 10^8 \frac{\eta}{\lambda(1-\eta)}~{\rm yr}
\end{equation}

\noindent as a proxy for accretion rate as a function of $\eta$, and plot $k$ vs $t_S$ to define an allowed region of parameter space in which quasars can (on average) lie (assuming no evolution in the mean $\eta$ or $\lambda$ and that the increase in the SMBH mass density is dominated by accretion). Here, a similar result is seen. With the GD07 local density we constrain $k$ to $<2$, with our derived density range then $0<k<3$ can (just) be accommodated (though only with very high values of $\eta \approx 0.55$ and $\lambda\approx 1$), but with the GWB local density range then an approximate lower limit can be set of $k>2$, assuming standard accretion disk dominated growth.

\section{Discussion}\label{sec:disc}

Conventional assumptions about SMBH accretion, combined with estimates of the local SMBH mass density based on the $M_{\bullet}-M_{\rm bulge}$ relation, imply an allowed range of $0<k\stackrel{<}{_{\sim}}2$ (Figure \ref{fig:salpeter}). This range is in slight tension ($\sim90\%$ confidence interval) with the results in \citet{2023ApJ...944L..31F}, though it is consistent with some of the recent constraints discussed in \S 1.2 \citep[e.g.][]{2023arXiv230503408L,rodriguez23, andrae23,2023OJAp....6E..25G}. 

Our constraints on $k$ from the Soltan argument are, however, dependent on four variables: the local SMBH mass density, the integrated luminosity density from the AGN luminosity function and the mean values of $\lambda$ and $\eta$, which regulate the total accretion derived from integrating the AGN luminosity function. We first briefly review the uncertainties on the luminosity density,  $\eta$ and $\lambda$ (the uncertainties on the local SMBH density are discussed in \S\ref{sec:bhdensity}), before discussing the effect of these uncertainties on our limit on $k$. 

\subsection{Uncertainties in the luminosity density}

The bolometric luminosity function of AGN is still uncertain, but most of the uncertainty lies at the faint end at high redshifts, whereas the dominant contribution to the luminosity density arises around the break in the luminosity function, where AGN surveys are relatively complete. 

As briefly discussed in \S \ref{sec:LF}, we adopt the luminosity function of \citet{2020MNRAS.495.3252S}, however, if, for example, we use the mid-infrared luminosity function of \citet{2018ApJ...861...37G} and convert to the bolometric luminosity function using the prescription of 
\citet{2015ApJ...802..102L} we recover a black hole mass density 60\% higher than that from the \citet{2020MNRAS.495.3252S} luminosity function. (Using this value would be even more constraining on the value of $k$.) Based on this small difference between two independently-derived luminosity density estimates we believe that the uncertainties from the AGN luminosity function, although not negligible, are smaller than those from the local black hole mass density.

\subsection{Uncertainties in accretion efficiency}
The typical accretion efficiency of an SMBH is poorly constrained \citep[e.g.][]{raimundo12}. The canonical value of $\eta\approx 0.1$ \citep[e.g.][]{1982MNRAS.200..115S} is not measured directly, but  obtained by matching early estimates of accretion luminosity to local SMBH mass density.
The theoretical limit is thought to be $\eta\approx0.3$ \citep{1974ApJ...191..507T} 
Observational estimates of $\eta$ vary from $\stackrel{<}{_{\sim}} 0.1$ to potentially $>0.5$ \citep[e.g.][]{trakhtenbrot14,jana22,farrah22,lai23smss}. Several factors are thought to affect $\eta$; for example, magnetized discs may have $\eta>0.3$ \citep{2016MNRAS.462..636A,Kinch_2021}, but there is no consensus on how these factors interrelate. For Figure \ref{fig:salpeter} we have assumed the conventional range of $0.05<\eta<0.3$, but a larger range is certainly not excluded. We note though that high values of $\eta$ reduce the available time for black hole growth by accretion in the early Universe, which is already problematic \citep[e.g.\ ][]{2018Natur.553..473B}.
 
\subsection{Uncertainties in Eddington ratio}
Directly observed Eddington ratios in quasars are consistent with $\lambda\sim0.05$, with a broad tail towards higher Eddington ratios \citep[e.g.][]{farrah23ellip}. 
This ratio may however be biased low. Most SMBH mass is likely
accreted in obscured phases \citep{martsan05} during which Eddington
ratios can be higher \citep{2014ApJ...785...19T,farrah22}. 
Furthermore, quasar surveys may find AGN whose SMBH masses and 
Eddington ratios are biased low and high, respectively \citep[e.g.][]{2008ApJ...680..169S}.

\section{Conclusions}\label{sec:conc}

By analysing the growth of the mass density in SMBHs via accretion we show that high values of cosmological coupling ($k\approx 3$) needed by \citet{2023ApJ...944L..31F} to account for dark energy via BHs is disfavored if conventional assumptions regarding the local SMBH mass density and accretion mechanisms are made. However, we cannot rule out scenarios in which $k=3$ is possible, and indeed, recent estimates of the GWB from pulsar timing arrays favor a much higher mass density of SMBHs than estimates based on the $M_{\bullet}-M_{\rm bulge}$ relation and would comfortably allow $k=3$ (Figure \ref{fig:salpeter}, right panel). Better constraints on the local mass function of SMBHs are thus essential both to our understanding of the GWB and for more reliably constraining any cosmological coupling of black holes.

\begin{acknowledgments}
We are grateful to L.Z.\ Kelley and the NANOGrav team for supplying the data from their simulations of supermassive black hole merger statistics. 
The National Radio Astronomy Observatory is a facility of the National Science Foundation operated under cooperative agreement by Associated Universities, Inc.
\end{acknowledgments}

\bibliography{bhgrowth}{}
\bibliographystyle{aasjournal}

\end{document}